\documentclass[fleqn,usenatbib]{mnras}

\usepackage{newtxtext,newtxmath}

\usepackage[T1]{fontenc}

\DeclareRobustCommand{\VAN}[3]{#2}
\let\VANthebibliography\thebibliography
\def\thebibliography{\DeclareRobustCommand{\VAN}[3]{##3}\VANthebibliography}

\usepackage{tabularx}

\usepackage{graphicx}	
\usepackage{amsmath}	

\title[Cold gas masses for spectroscopic surveys]{Cold gas mass measurements for the era of large optical spectroscopic surveys}

\author[Scholte \& Saintonge]{
Dirk Scholte$^{1}$\thanks{E-mail: dirk.scholte.20@ucl.ac.uk} and 
Am\'elie Saintonge$^{1}$
\\
$^{1}$Department of Physics and Astronomy, University College London, London, WC1E 6BT, UK
}

\date{Accepted XXX. Received YYY; in original form ZZZ}

\pubyear{2022}

\begin{document}
\label{firstpage}
\pagerange{\pageref{firstpage}--\pageref{lastpage}}
\maketitle

\begin{abstract}
Gas plays an important role in many processes in galaxy formation and evolution, but quantifying the importance of gas has been hindered by the challenge to measure gas masses for large samples of galaxies. Datasets of direct atomic and molecular gas measurements are sufficient to establish simple scaling relations, but often not large enough to quantify three-parameter relations, or second order dependencies.  As an alternative approach, we derive here indirect cold gas measurements from optical emission lines using photoionization models for galaxies in the SDSS main galaxy sample and the PHANGS-MUSE survey. We calibrate the gas surface density measurements using xCOLD GASS and PHANGS-ALMA molecular gas measurements to ensure our measurements are reliable. We demonstrate the importance of taking into account the scale-dependence of the relation between optical depth ($\tau_V$) and gas surface density ($\Sigma_{\textrm{gas}}$) and provide a general prescription to estimate $\Sigma_{\textrm{gas}}$ from $\tau_V$, metallicity and the dust-to-metal ratio, at any arbitrary physical resolution. To demonstrate that the indirect cold gas masses are accurate enough to quantify the role of gas in galaxy evolution, we study the mass-metallicity relation (MZR) of SDSS galaxies and show that as a third parameter, gas mass is better than SFR at reducing the scatter of the relation, as predicted by models and simulations. 
\end{abstract}




\section{Introduction}
Gas is one of the most important ingredients in describing the formation and evolution of galaxies \citep{tinsley1980}. It is the fuel for star formation and an important contributor to the total mass of galaxies. The cold interstellar medium is sensitive to the properties of the circumgalactic medium and the large-scale environment, which regulate its total mass through accretion, and on the other hand its properties determine the rate and efficiency of the star formation process \citep[e.g.][]{saintonge22}. To study these multi-scale processes in detail, reliable measurements of the cold gas content of large samples of galaxies are required.

There have been extensive efforts to derive atomic and molecular gas masses for galaxies in the local universe. The total atomic gas mass of galaxies in the local universe has been measured for large galaxy samples through observations of the HI 21cm emission line as part of both blind and targeted surveys such as HIPASS \citep{barnes2001}, HIJASS \citep{lang2003}, EBHIS \citep{winkel2010}, ALFALFA \citep{haynes2011} and xGASS surveys \citep{catinella2018}. The resolved atomic gas surface densities have been studied through surveys such as THINGS \citep{walter2008} and VIVA \citep{chung2009}.  Molecular gas masses are most often measured via observations of emission lines of the CO molecule, in particular CO(1-0) and CO(2-1) in the nearby universe, extrapolated to total molecular masses via the CO-to-H$_2$ conversion function, $\alpha_{CO}$ . This technique was used to derive the integrated gas content of local galaxies in for example the FCRAO \citep{young95}, xCOLD GASS \citep{saintonge2017}, and  ALLSMOG \citep{cicone17} surveys. Other surveys have mapped CO at kiloparsec-scales for samples of $\sim$50-100 galaxies [e.g. HERACLES \citep{leroy2013}, EDGE-CALIFA \citep{bolatto2017}, ALMaQUEST \citep{lin2019}], and at 100s-pc scales \citep[PHANGS-ALMA, ][]{leroy2021}. These surveys established our understanding of the atomic and molecular gas content and distribution in galaxies, as well as the cosmic cold gas mass density in the local Universe \citep[e.g.][]{casasola2017, jones2018, fletcher2021}. General trends in the cold gas content and distribution are quantified as a function of galaxy type, stellar mass, star formation rate and other parameters through a range of scaling relations \citep[see references above, and overview in ][]{saintonge22}. These measurements have driven greater understanding of the role of cold gas in galaxy evolution, and provide tests on  models and simulations \citep[e.g.][]{lilly2013, dave2020, feldmann2020}. 

Important questions however remain unanswered, as even larger samples are required to further our understanding of the role of cold gas in processes in galaxy physics, especially those relating to environment, large scale structure, and dark matter halo properties. To study the complicated ways gas regulates star formation and galaxy evolution, we need samples several orders of magnitude larger than can currently be derived through direct observations of CO or HI emission lines. However, such large samples of gas mass measurements can be obtained through indirect methods that rely observations of dust in either emission or absorption. These methods assume dust and gas are well mixed, which allows the conversion of dust mass to gas mass through a dust-to-gas ratio. Dust masses are most accurately measured through observation of dust emission at far infrared wavelengths that probe the peak or Rayleigh-Jeans tail of the thermal emission \citep[e.g.][]{scoville2014}. Cold gas mass estimates from dust emission produce measurements with a scatter of $\sim$0.15 dex \citep{janowiecki2018}.

An alternative is to estimate the gas mass from optical spectroscopy, relying on the measurement of dust attenuation through the Hydrogen Balmer lines, in combination with metallicity estimates to estimate the dust-to-gas ratio in the ISM of galaxies \citep{guver2009, heiderman2010, brinchmann2013}. Gas surface densities derived from the Balmer decrement trace the total gas content along the line of sight instead of atomic or molecular gas separately \citep{concas2019, barrera-ballesteros2020}. These measurements  correlate with direct measurements of atomic and molecular gas \citep{concas2019, yesuf2019, piotrowska2020, barrera-ballesteros2020}. However, the scatter in these gas surface density measurements is large compared to other methods at $\sim$0.3 dex for global observations and larger scatter for resolved observations. The calibration of these measurements to direct atomic and molecular gas measurements is difficult due to the different physical regions and gas phases traced. Differences in aperture sizes and observational resolutions further complicate the calibration of gas mass measurements through optical spectroscopy. 

Despite these complications, this technique has the potential of providing the largest samples of indirect cold gas masses through surveys such as SDSS \citep[][]{abazajian2009, piotrowska2020}. Future surveys will further expand this sample in the coming years. Most imminently, the DESI survey will push the number of galaxies with optical spectra into the tens of millions \citep{desi2022}. This will be followed by the 4MOST, WEAVE and MOONS surveys \citep[][]{driver2019, dalton2012, cirasuolo2020}. Together these surveys will expand the number of galaxy spectra by several orders of magnitude and open up parameter space that has so far remained out of reach (e.g. higher redshifts and lower stellar masses). This context highlights the potential of spectroscopic cold gas mass measurements, if reliable calibrtions can be derived.  

In this study, we derive gas surface densities and gas masses from SDSS and PHANGS-MUSE spectroscopy using photoionization models and calibrate our measurements with integrated observations from the xCOLD GASS survey and resolved observations from the PHANGS-ALMA survey \citep[][]{emsellem2021, leroy2021}. We apply our gas mass measurements to study the role of gas in the scatter of the relation between stellar mass and gas phase metallicity.

In Section \ref{sec:photoionization_model} we describe the photoionization models and the simulation based inference methods. In Section \ref{sec:model_validation} we validate the measurements we derive from our models and in Section \ref{sec:gas_surface_density} we calibrate the gas surface densities we derive using PHANGS-MUSE and xCOLD GASS. In Section \ref{sec:MZR} we show how the gas mass measurements we derive can be used to measure the role of gas in the MZR. We summarize our results and draw conclusions in Section \ref{sec:conclusion}.

\section{Photoionization model}
\label{sec:photoionization_model}
Photoionization models allow us to simulate the optical line emission of star forming galaxies with a large variety of physical conditions. These simulated line emissions are derived from basic physical principles such as radiative transfer, chemical gas composition, dust depletion and dust attenuation, under the assumption of a spherical shell geometry for the gas with a central ionisation source. When the line emissions of observed galaxies are matched to these simulations we unlock important information about the physical conditions of the ISM in galaxies. We construct our models with five free parameters: unattenuated H-alpha flux ($F_{H\alpha}$), total metallicity ($Z$), ionization parameter ($U$), dust-to-metal ratio ($\xi$) and optical depth ($\tau_V$). The photoionization models are used to train a neural network for posterior inference using Simulation Based Inference \citep{tejero-cantero2020}. The trained neural network is then applied using measured emission line fluxes to infer the best values of our five free parameters.

We use a set of 8 emission lines to recover the free parameters in our photoionization models: [OII]3727\AA, [OII]3729\AA, H$\beta$4861\AA, [OIII]4959\AA, [OIII]5007\AA, [NII]6548\AA, H$\alpha$6563\AA, and [NII]6584\AA. This set of emission lines contains the majority of strong emission lines in the optical spectra of star forming galaxies. The [SII]6717\AA\ and [SII]6731\AA emission lines are excluded from our procedure. Whilst we could include these lines in our measurements we have chosen not to as there are no unambiguous measurements of the amount of sulfur locked in dust grains \citep{jenkins2009}, which can have a significant impact on the [SII]6717 and [SII]6731 emission line fluxes. Additionally, the wavelength of the Sulfur optical strong lines means they are redshifted outside of the wavelength range of surveys such as SDSS and DESI for all but the lowest redshifts. 

\subsection{Model description}
We use the Cloudy photoionization code, version C17.02 \citep{ferland2017} and the pyCloudy Python wrapper for Cloudy \citep{morisset2013}.  The photoionization models are produced using similar procedures as in \cite{charlot2001, brinchmann2013} and \cite{byler2017}. Generally, we implement the same parametrization as used in \cite{brinchmann2013} but implement improvements to the photoionization models introduced by \cite{byler2017}. 

We use FSPS to produce synthetic stellar population spectra \citep{conroy2009,conroy2010} using MIST isochrones \citep{choi2016}. These are produced using a Chabrier IMF \citep{chabrier2003} with a lower mass limit of 0.08 $M_\odot$ and an upper mass limit of 120 $M_\odot$. We adopt solar abundances from \cite{grevesse2010} and depletion factors from \cite{dopita2013}. The metallicity scaling of secondary nuclear synthesis elements is implemented using \cite{vanzee1998}. The irradiating spectrum we use is based on 2 Myr of star formation using a constant star formation history. The assumed age of 2 Myr is within the range of ages where stars produce ionizing radiation consistent with HII regions in BPT diagrams \citep{byler2017}. The metallicities of the single stellar population spectra are sampled more coarsely than the gas phase metallicities. Therefore we use the SSP metallicity that matches best the undepleted gas phase metallicity.

The assumptions made in these photoionization models aim to reproduce the properties of typical HII regions, rather than the full possible range exhibited across an entire galaxy. This needs to be taken into account when interpreting the results derived from the models. The constraints derived from applying the models to a single HII region may not accurately describe its physical state. However, when applied to large samples of HII regions, we can use the  photoionization models to study the average physical conditions of the ISM. Naturally, similar caveats are applicable to other calibrations of strong line metallicity estimates.

\subsubsection{Geometry and ionizing spectrum intensity}
The models assume a central irradiating source within a sphere of gas of uniform density, as in \cite{byler2017}. We assume a Hydrogen density ($n_H$) of 100 cm$^{-3}$ and an inner radius of the gas cloud ($R$) of $10^{19}$ cm ($\sim 3$ pc). 

In these models the rate of Hydrogen ionizing photons of a single stellar population (SSP) is defined as:
\begin{equation}
    \hat{Q}(t) \equiv \frac{1}{hc} \int_0^{\lambda_L} d\lambda \lambda S_{\lambda}(t) ,
\end{equation}
where $S_{\lambda}(t)$ is the luminosity emitted per unit wavelength per solar mass by a stellar generation of age $t$, $\lambda_{L}$ is the maximum wavelength capable of ionizing Hydrogen atoms (912 \AA) and $c$ is the speed of light. The rate of ionizing photons is set through the ionization parameter. This parameter is defined as:
\begin{equation}
    U \equiv \frac{Q(t)}{4 \pi R^2 n_H c},
\end{equation}
where $Q(t)$ is the total rate of ionizing photons of the ionizing source. The ionization parameter value dictates the luminosity and stellar mass needed to produce the number of ionizing photons with the physical conditions specified as $Q(t) = M_* \times \hat{Q}(t)$. 

Following the stopping criteria of the photoionization models by \cite{byler2017} we set a temperature floor of 100K and stop the radiative transfer calculation when the ionized fraction of the cloud drops to 1\%.

\subsubsection{Chemical abundances}
The cosmic abundances used in our model are based on solar abundances from \cite{grevesse2010}, where $Z_{\odot}$ = 0.0142. The solar abundances of the elements implemented in our model are shown in Table \ref{tab:metallicities_depletions}. The elemental abundances are scaled linearly with metallicity $Z/Z_{\odot}$, except for Helium, Carbon and Nitrogen. For these elements, we adopt the scalings of \cite{dopita2013}, which implement the measured scaling of Carbon and Nitrogen as a function of Oxygen abundance of \cite{vanzee1998}, accounting for the production of these elements through secondary nucleosynthesis.

\begin{table}
 \centering
 \caption{Solar abundances and depletion factors as adopted in the models. Solar abundances from \protect \cite{grevesse2010} and reference depletion factors from \protect \cite{dopita2013}.}
 \begin{tabularx}{\columnwidth}{p{0.3\columnwidth} p{0.3\columnwidth} p{0.3\columnwidth} }
\hline
\hline 
\textbf{Element (X)} & \textbf{log(X/H)} & \textbf{reference log($D_{\textrm{gas}}$})             \\ \hline
H                    & 0                 & 0                                       \\
He                   & --1.01            & 0                                       \\
C                    & --3.57            & --0.30                                  \\
N                    & --4.60            & --0.05                                  \\
O                    & --3.31            & --0.07.                                 \\
Ne                   & --4.07            & 0                                       \\
Na                   & --5.75            & --1.00                                  \\
Mg                   & --4.40            & --1.08                                  \\
Al                   & --5.55            & --1.39                                  \\
Si                   & --4.49            & --0.81                                  \\
S                    & --4.86            & 0                                       \\
Cl                   & --6.63            & --1.00                                  \\
Ar                   & --5.60            & 0                                       \\
Ca                   & --5.66            & --2.52                                  \\
Fe                   & --4.50            & --1.31                                  \\
Ni                   & --5.78            & --2.00                                  \\ \hline
\end{tabularx}
\label{tab:metallicities_depletions}
\end{table}

Our models differ from those of \cite{byler2017} as we introduce a variable dust-to-metal ratio ($\xi$), as in the \cite{charlot2001} and \cite{brinchmann2013} models. The dust-to-metal ratio is defined as the mass ratio ($M_{d}/M_{Z}$) between the total dust mass and the total mass of metals both in the gas and dust phase, where $M_{Z}$ is the mass sum of all elements heavier than Helium. We vary the dust-to-metal ratio through a linear scaling with log($D_{\textrm{gas}}$) for each element,  similar to the method of \cite{jenkins2009}. For gas with $Z = Z_{\odot}$  the depletion factor $D_{\textrm{gas}}$ of element X is defined as 
\begin{equation}
    \textrm{log}(D_{\textrm{gas}})=\textrm{log}\{N(X)/N(H)\} - \textrm{log}\{N(X)/N(H)\}_{\odot}.
\end{equation}

We anchor the scaling of the depletion factors of individual elements at zero depletion and at the depletion values of \cite{dopita2013}, which can be found in Table \ref{tab:metallicities_depletions}. From these anchor points, we make linear extrapolations in log($D_{\textrm{gas}}$) for each element. 

\subsubsection{Optical depth and gas surface density}
We calculate the optical depth of the ISM following \cite{charlot2000}. This defines a transmission function:
\begin{equation}
\label{CH2000_transmission}
    T_{\lambda} = e^{-\tau_{\lambda}}
\end{equation}
\begin{equation}
\label{dust_attenuation}
    \tau_{\lambda} = \tau_{V} \left[ \frac{\lambda}{5500 
\mathring{\mathrm{A}}} \right]^{-n}
\end{equation}

We use an exponent $n = 1.3$, this value being representative of the optical depth for young stars in birth clouds \citep{dacunha2008}. Gas surface densities can be derived using the optical depth measurement and other free parameters in our model. They are calculated through the following equation derived by \cite{brinchmann2013}: 
\begin{equation}
    \label{eq:gas_equation_brinchmann}
    \Sigma_{\textrm{gas}} = 0.2 \frac{\tau_{V}}{\xi Z} (M_{\odot}\textrm{ pc}^{-2})
\end{equation}
where the combination of $\xi Z$ is the gas-to-dust ratio.
\begin{table}
 \centering
 \caption{Ranges of the uniform priors on the model parameters. The steps are the number of photoionization models generated along the direction of each parameter that was modelled with Cloudy.}
 \label{tab:model_priors}
\begin{tabularx}{\columnwidth}{lll}
\hline
\hline
\textbf{Param.} & \textbf{Description}            & \textbf{Range}                                \\ \hline
$F_{H\alpha}$      & Unattenuated H$\alpha$ flux     & --17 $\leq$ log($F_{H\alpha}/$erg cm$^{-2}$ s$^{-1}$) $\leq$ --13       \\
$Z$                & Total metallicity               & --1.0 $\leq$ log($Z/Z_{\odot}$) $\leq$ 0.7 (24 steps)   \\
$U$                & Ionizaton parameter             & --4.0 $\leq$ log($U$) $\leq$ --1.0 (33 steps) \\
$\xi$              & Dust-to-metal ratio             & 0.1 $\leq$ $\xi$ $\leq$ 0.6 (9 steps)         \\
$\tau_{V}$         & Total optical depth             & --2 $\leq$ log($\tau_{V}$) $\leq$ 0.6             \\
\hline
\end{tabularx}
\end{table}

\begin{figure*}
    \centering
    \includegraphics[width=\textwidth]{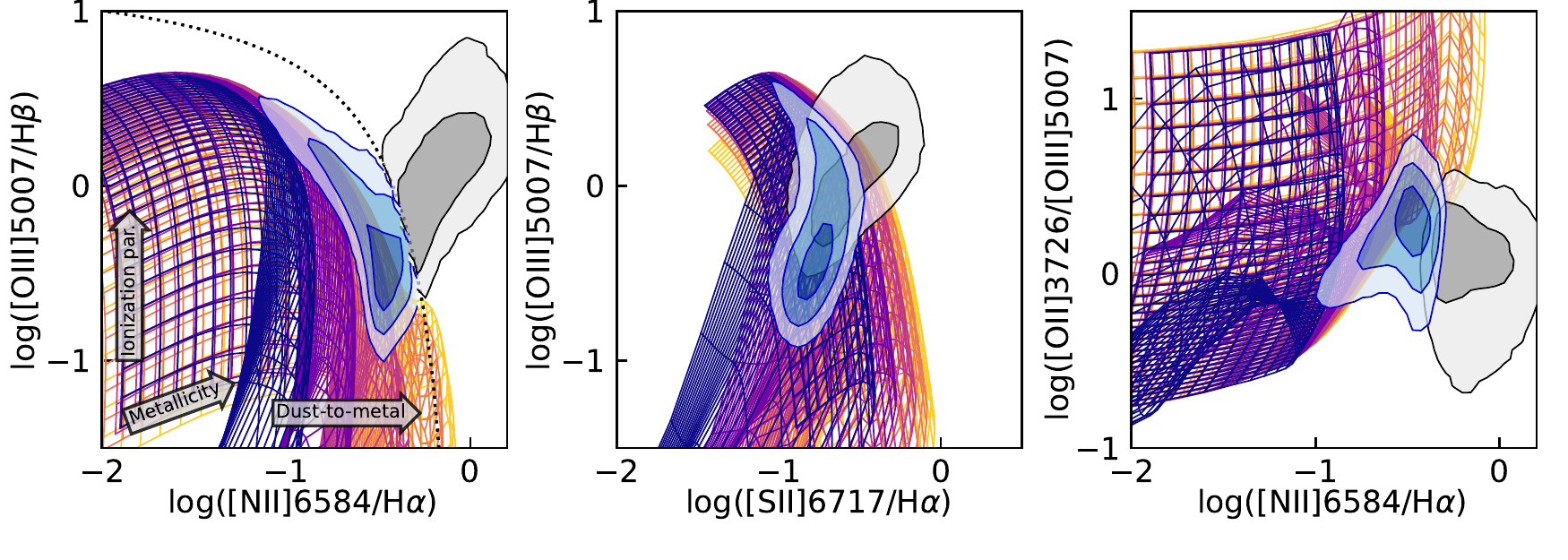}
    \caption{Modelled emission line ratios plotted onto BPT diagrams of SDSS DR7 galaxies (star forming (blue) and not star forming (grey) \citep{kauffmann2003}). The modelled total metallicity increases along the horizontal lines on the left side of the left plot, where -1.0 < $Z$ < 0.7. The vertical lines in the same region are the model evaluated at various effective ionization parameters, where -4.0 < $U$ < 0.0). The separate grids with different colours are models evaluated for different depletion factors, where 0.1 < $\xi$ < 0.6 (from low to high: dark to light). The Sulfur emission lines are not used in our analysis (see Section \ref{sec:photoionization_model}) but are predicted by the photoionization models. Here the models are shown for unattenuated emission. Once dust attenuation and noise are taken into account our models cover the full parameter space covered by the SDSS star forming galaxies.}
    \label{fig:photoionization_model_BPT}
\end{figure*}

\subsection{Simulation based inference}
We use simulation based inference as it allows us to perform parameter inference at a fraction of the computational cost of standard inference procedures using MCMC sampling. This makes it possible to apply our model to the largest survey samples available. We use the emission line fluxes produced by the photoionization model over a large parameter space in $F_{H\alpha}$, $Z$, $U$, $\xi$ and $\tau_V$ to train a sequential neural posterior estimation model (SNPE, \cite{greenberg2019}) using masked amortized flow.  We draw $5\times10^6$ samples from the uniform prior distributions defined in Table \ref{tab:model_priors}. Using the drawn parameters we simulate emission line fluxes using the photoionization model, and a Gaussian noise model. The SNPE model is trained using the standard settings in the Simulation Based Inference package \citep{tejero-cantero2020}. This includes a learning rate of $5 \times 10^{-4}$ using an Adam optimizer, a validation fraction of 0.1 and a training stopping criterion of 20 epochs without improvement on the validation set. Because it does not generate significant additional computational cost, Simulation Based Inference makes it possible to include $F_{H\alpha}$ as a free parameter, instead of scaling all the measurements by the H$\alpha$ flux. This provides a straightforward way to take into account the uncertainties in the H$\alpha$ flux measurement. 

The choice of prior is of particular importance for the $\xi$ parameter as it is only weakly constrained by the emission line measurements. The effect of the priors on the posterior distributions was explored by \cite{brinchmann2013}, therefore, we adopt the same prior on $\xi$. We adopt uniform priors in log-space on the other parameters. However, this is of lesser influence as these parameters are well constrained by the emission line measurements.

Due to the computational demands of the photoionization modelling we do not run a Cloudy model for each drawn sample during training of the neural network. Instead we produce a grid of $24 \times 33 \times 9 = 7128$ photoionization models for parameters log($Z$), log($U$) and $\xi$. The dense sampling allows us to approximate the continuous relation between our parameters and emission line fluxes through a linear interpolation of the modelled grid points. The dust attenuation correction is applied for each drawn sample directly using the optical depth calculated. The Cloudy models predict emission line fluxes for a large set of emission lines. The results of some of the most important emission line flux predictions are shown in figure \ref{fig:photoionization_model_BPT}. These are BPT diagrams \citep{baldwin1981} showing the modelled emission line strengths in grids of metallicity and ionization parameter. The different colours represent models with different dust-to-metal ratios. The blue and black contours overplotted are the measured emission line flux ratios for SDSS star-forming galaxies and AGN, respectively \citep{brinchmann2004, kauffmann2003, tremonti2004}. The models show line ratios for unattenuated emission.

We use an additive Gaussian noise model to simulate noise in the training sample. The noise model is dependent on the observed wavelength which allows us to take into account the variable noise levels due to the wavelength dependence on the spectrograph throughput. We use the noise distribution in the used data set to simulate the noise in the training sample. We use the full array of emission line flux measurements and emission line flux errors as the input vector for the inference model. We input zero for the flux and flux error when the wavelength of an emission line is outside the wavelength range of the spectrograph. We also apply this method to the simulated data to train the inference procedure on missing data. 

\begin{figure}
    \centering
    \includegraphics[width=\columnwidth]{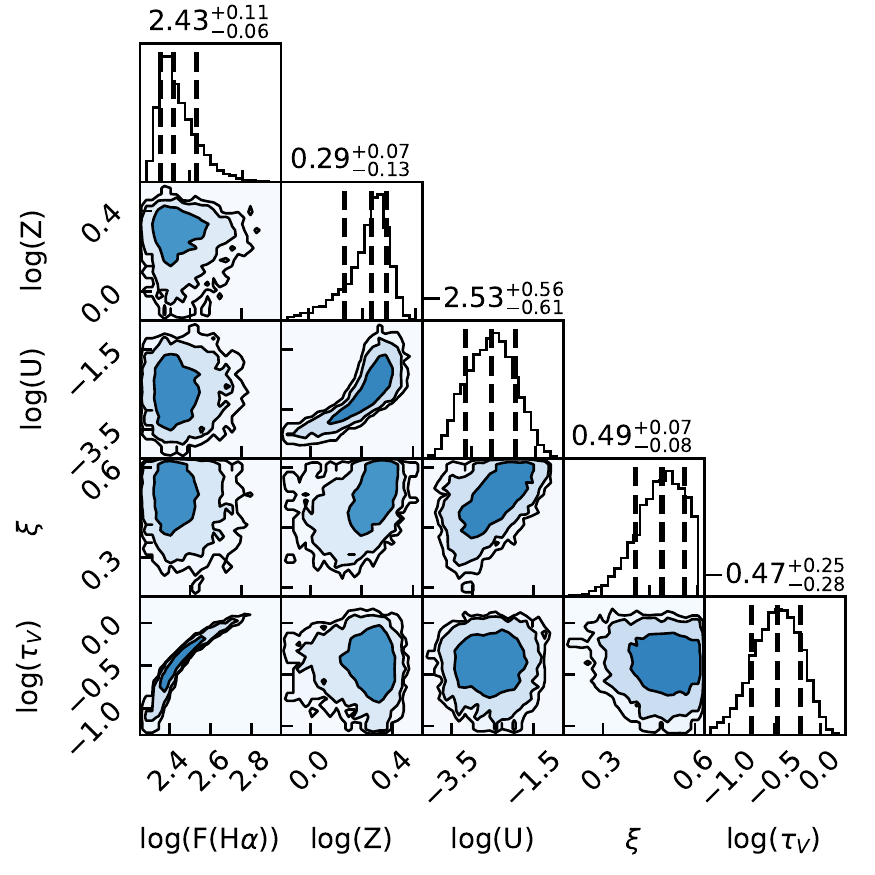}
    \caption{An example of a posterior distribution derived using Simulation Based Inference. This posterior was derived for a main sequence galaxy in our sample with SPECOBJID 492246374255951872 in SDSS DR8. This galaxy is also part of the xCOLD GASS sample. The red star on Figure \ref{fig:MH2_FH2_SFE_comparison_xCOLDGASS} marks this galaxy on the $M_{*}$-SFR plane.}
    \label{fig:corner_plot_example}
\end{figure}

In Figure \ref{fig:corner_plot_example} we show the posterior distribution of one of the galaxies in our sample. The contours show the 1, 2 and 3-$\sigma$ ranges from dark to light. This galaxy is also part of the xCOLD GASS sample and its location on the $M_{*}$-SFR plane is marked by the red star on Figure \ref{fig:MH2_FH2_SFE_comparison_xCOLDGASS}. This figure clearly shows the degeneracy between some of the parameters in the photoionization models. Most notably there is a clear degeneracy between metallicity and the ionization parameter and between unattenuated H$\alpha$ flux and optical depth. It is clear from this figure that the inference procedure we use is able to explore these degenerate non-Gaussian parameter spaces, as has also been shown for other inference problems \citep{greenberg2019}.

\section{Model validation}
\label{sec:model_validation}
We validate the results of our models by comparing the derived parameters to either direct measurements of the same quantities, or estimates using other methods. We use emission line flux measurements from the SDSS DR8 MPA-JHU catalogue \citep{brinchmann2004, kauffmann2003a, tremonti2004}. From this catalogue we select star forming galaxies using the criterion described in \cite{kauffmann2003}. We apply a signal-to-noise selection of 15$\sigma$ on the H$\alpha$ emission line only, to ensure we only include high quality measurements whilst avoiding selection biases which could arise if additional signal-to-noise criteria were applied to other emission lines \citep{yates2012, kashino2016}. We select galaxies with redshifts with $z \geq 0.027$. Within this range we have access to the [OII]3727 and [OII]3729 emission lines, which are important to constrain the full range of parameters we use in our photoionization models. For each parameter measurement we use the 50\textsuperscript{th} percentiles of the marginalized posteriors to determine the measured value, and use the 16\textsuperscript{th} and 84\textsuperscript{th} percentile measurements to define the uncertainties.

In the inference procedure using the SDSS MPA-JHU emission line flux measurements we multiply the formal emission line flux errors by the recommended multiplication factors\footnote{https://wwwmpa.mpa-garching.mpg.de/SDSS/DR7/raw\_data.html} to account for additional uncertainties due to continuum subtraction. Our signal-to-noise selections are made using the formal flux errors.

\subsection{In-fiber star formation rates}
\begin{figure}
    \centering
    \includegraphics[width=\columnwidth]{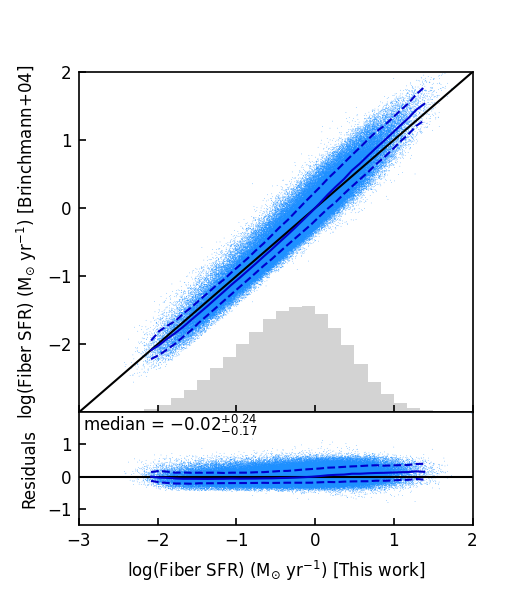}
    \caption{A comparison between the fiber star formation rates derived and the fiber star formation rates derived by \protect \cite{brinchmann2004} for star forming galaxies. A histogram of the distribution of star formation rates we derive is shown in grey.}
    \label{fig:model_SFR_vs_brinchmann04}
\end{figure}

We derive fiber star formation rates (SFRs) from the inferred unattenuated H$\alpha$ flux. We convert the flux to luminosity using the redshift derived luminosity distance. The H$\alpha$ luminosity is converted to SFR using \cite{kennicutt1998}:
\begin{equation}
    \textrm{SFR} \approx 7.9 \times 10^{-42} \left( \frac{L_{H\alpha}}{\textrm{erg s}^{-1}} \right) (\textrm{M}_{\odot} \textrm{ yr}^{-1}) .
\end{equation}
The derived SFRs are compared to the in-fiber SFRs from \cite{brinchmann2004} in Figure \ref{fig:model_SFR_vs_brinchmann04}. These SFR measurements are derived through a similar method to ours, and as expected, the two sets of measurements are in agreement with a median offset of $-$0.05$_{-0.15}^{+0.24}$. The weak trend in the residuals can be explained by the different attenuation curve used in \cite{brinchmann2004} compared to this work.

\subsection{Gas phase metallicity}
\begin{figure*}
    \centering
    \includegraphics[width=\textwidth]{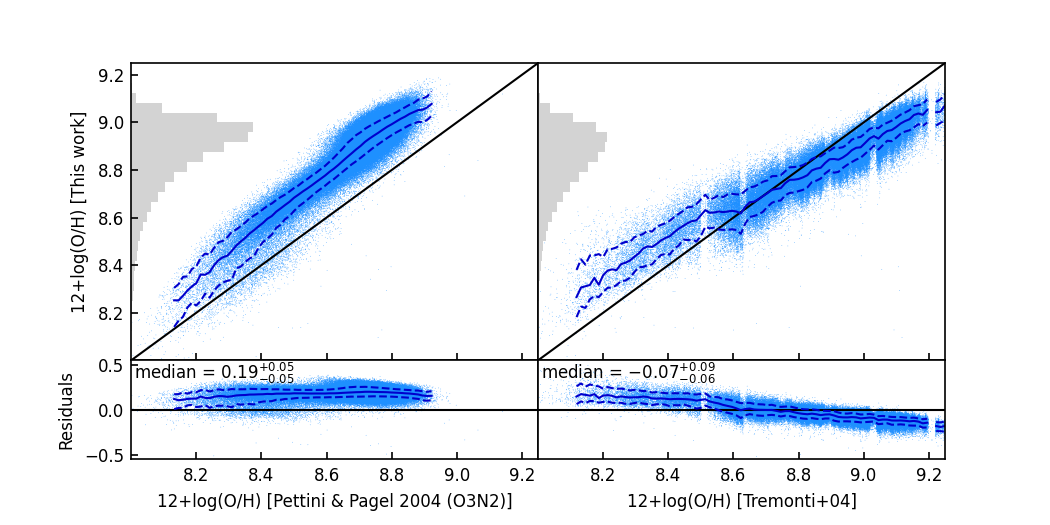}
    \caption{A direct comparison between the gas phase metallicities we derived and metallicities derived using the \protect \cite{pettini2004} O3N2 estimator (left) and \protect \cite{tremonti2004} metallicities (right). The one-to-one relation (black), median offset (red), and running median (blue) are shown in the overplotted lines. A histogram of the distribution of the metallicities we derive is shown on the left of each plot. The plot on the right contains less data points as we show only the cross-matched sample with \protect \cite{tremonti2004} metallicities.}
    \label{fig:model_metallicity_vs_PP04_O3N2}
\end{figure*}

We compare our gas phase metallicities to metallicities derived using (1) the \cite{pettini2004} calibration of the O3N2 metallicity indicator, and (2) the photoionization modelling of \cite{tremonti2004}. The O3N2 indicator is given by:
\begin{equation}
\label{eq:O3N2_indicator}
    \textrm{O3N2 = log}\left(\frac{\textrm{[OIII]}5007 / \textrm{H}\beta}{\textrm{[NII]}6584 / \textrm{H}\alpha}\right)
\end{equation}
as described by \citet{alloin1979}, and the metallicity calibration from \cite{pettini2004} is:
\begin{equation}
    \textrm{12 + log(O/H)} = 8.73 - 0.32 \times \textrm{O3N2}.
\end{equation}
Figure \ref{fig:model_metallicity_vs_PP04_O3N2} shows the one-to-one relation (black), a running median (dark blue, 16$^{\textrm{th}}$ and 84$^{\textrm{th}}$ percentiles dashed) and the individual measurements (light blue). There is a median offset of 0.18$_{-0.06}^{+0.05}$ between the metallicities using the \cite{pettini2004} calibration and our estimates. This is in agreement with expectations, as theoretical metallicity derivations using photoionization modelling typically result in higher values than empirical calibrations \citep{kewley2008}. Our comparison with the \cite{tremonti2004} metallicities show a smaller median offset (-0.08$_{-0.06}^{+0.08}$) but an overall trend in the median offset as a function of metallicity. Overall, barring a constant offset we get more consistent agreement with the empirical O3N2 metallicity calibration from \cite{pettini2004} than with the theoretical calibration from \cite{tremonti2004} for gas phase metallicities. The discrepancy with the metallicity measurements from \cite{tremonti2004} can be explained by differences in the photoionization models such as the implementation of the metallicity scaling of secondary nucleosynthesis elements (Carbon, Nitrogen and Oxygen) and elemental depletion factors. We also compare the metallicities returned by our model to a larger set of calibrations in Appendix \ref{sec:appendix_metallicities}. These results show that our results are within the range of metallicities derived through both theoretical and empirical calibrations available. 

\subsection{Optical depth}
\begin{figure}
    \centering
    \includegraphics[width=\columnwidth]{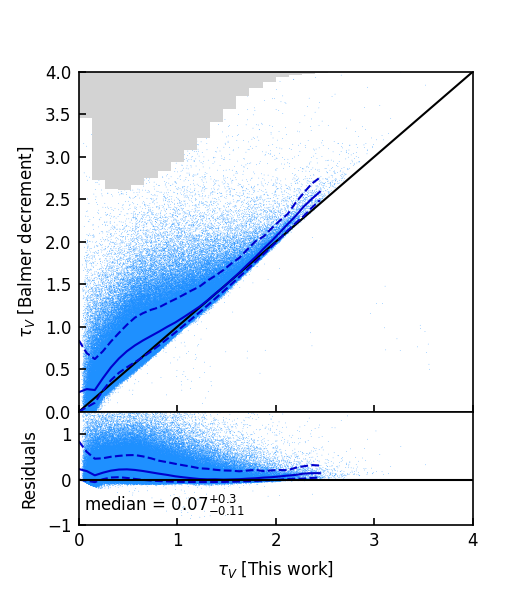}
    \caption{A comparison between the optical depth values we derived and optical depth derived using the Balmer decrement. A histogram of the distribution of optical depths we derive is shown in grey.}
    \label{fig:model_dust_attenuation_vs_Balmer_decrement}
\end{figure}

We compare the optical depth measurements we derive using our photoionization modelling and inference procedure to optical depths derived just from the Balmer decrement in Figure \ref{fig:model_dust_attenuation_vs_Balmer_decrement}. The optical depths directly derived from the Balmer decrement are calculated through the same method we use to attenuate our photoionization model fluxes:
\begin{equation}
\label{eq:optical_depth_BD}
    \tau_{V} = \frac{\lambda_{V}^{-n}}{\lambda_{\textrm{H}\beta}^{-n} - \lambda_{\textrm{H}\alpha}^{-n}} \textrm{ln}\left( \frac{\textrm{H}\alpha/\textrm{H}\beta}{2.86} \right) .
\end{equation}

Our measurements are in agreement with the optical depths derived using the Balmer decrement, with a small offset of $-$0.07$_{-0.11}^{+0.3}$. For some of the low signal-to-noise measurements we derive slightly lower median optical depths as the data is unable to fully constrain the optical depth measurement. The overall agreement shows that the optical depth we derive through our modelling is mostly determined by the H$\alpha$ and H$\beta$ flux measurements. The addition of the other emission lines does not significantly influence the measurement. This is an expected outcome, as the emission line flux of the other lines such as [NII] and [OIII] are dependent on the other free parameters in our model, and therefore much less constraining than the Hydrogen lines in the specific case of attenuation. 

\begin{figure*}
    \centering
    \includegraphics[width=0.8\textwidth]{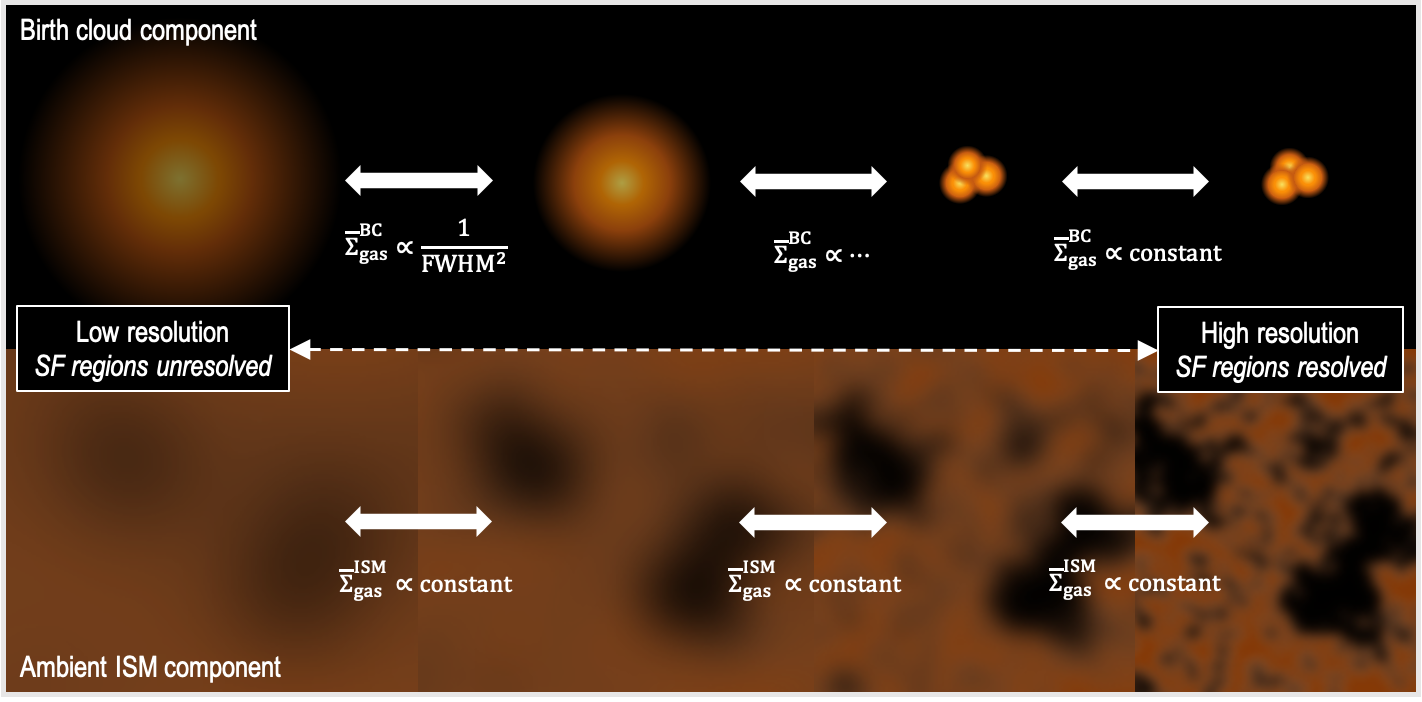}
    \caption{A diagram showing the expected resolution dependence of the birth cloud component ($\Sigma_{\textrm{gas}}^{\textrm{BC}}$, top panel) and ambient ISM component ($\Sigma_{\textrm{gas}}^{\textrm{ISM}}$, bottom panel) at different resolutions. In the unresolved regime the measured gas surface density of a birth cloud scales with FWHM$^{-2}$ due to the beam dilution of the CO(2-1) flux over larger areas at low resolutions. Once the birth cloud is resolved there no longer is a resolution dependence. In practice simple relation is complicated by the effects of multiple birth clouds are taken into account. The resolution at which the transition between the different regimes occurs is dependent on the size of gas reservoirs in a galaxy. In contrast, the average measured surface density of the ambient ISM component is resolution independent.}
    \label{fig:diagram_resolution_dependence_gas_cloud}
\end{figure*}

\section{Applications}

\subsection{Optical depth as a proxy for gas surface density}
\label{sec:gas_surface_density}
Optical depth is an often used proxy to estimate gas surface density and gas mass of galaxies \citep[e.g.][]{guver2009, heiderman2010, brinchmann2013, concas2019}. We use MUSE observations from the PHANGS-MUSE survey \citep{emsellem2021} to study the resolved relation between optical depth and gas surface density through comparison to the PHANGS-ALMA survey molecular gas surface densities \citep{leroy2021}. We also compare cold ISM gas surface densities estimated from SDSS emission line flux data using our photoionization model to direct CO measurements from the xCOLD GASS survey \citep{saintonge2017}.

\subsubsection{Resolved gas surface density comparisons to PHANGS-ALMA}
As discussed in the Introduction, the relation between optical depth and gas surface density is a useful proxy to estimate the atomic and/or molecular gas content of large samples of galaxies. While the general method can also be applied to mapping  observations, significant care must be taken because globally validated relations between optical depth and gas surface density (e.g.  Eq. \ref{eq:gas_equation_brinchmann}), cannot be applied directly to resolved observations.

There are two components that contribute to the dust attenuation of optical emission lines and derived gas surface densities. The first component (1) is the dust attenuation imparted by the dust present in the birth cloud, which is physically associated with the star forming region where the optical line emission originates, ($\Sigma_{\textrm{gas}}^{\textrm{BC}}$). The second component (2) is the dust attenuation due to other dust along the line of sight in the ambient ISM ($\Sigma_{\textrm{gas}}^{\textrm{ISM}}$), still within the galaxy but not directly associated with the star forming region where the emission originates. The total attenuation is the attenuation due to the sum of the two dust/gas components \citep[][]{calzetti1994, calzetti2000, charlot2000}.

Of these two components, the first one ($\Sigma_{\textrm{gas}}^{\textrm{BC}}$) produces a resolution dependence in the gas surface density estimates based on optical depths because it is physically associated with the immediate source of the radiation. The second component ($\Sigma_{\textrm{gas}}^{\textrm{ISM}}$) does not have this resolution dependence. In galaxies with high optical depth, $\Sigma_{\textrm{gas}}^{\textrm{BC}}$ has a large impact on the overall dust attenuation as there are many gas/dust rich star forming regions still associated with their birth clouds. In galaxies with low optical depth, we expect the foreground term, $\Sigma_{\textrm{gas}}^{\textrm{ISM}}$, to be the dominant. For this reason, the global relation between optical depth and gas surface density is a good approximation even for resolved observations when optical depths are low. 

The resolution dependence of the measured gas surface density of a birth cloud is expected to scale with FWHM$^{-2}$ until the birth cloud (or cluster of birth clouds) is resolved; this is due to the beam dilution of the flux at low resolutions. Once the resolution reaches the scales where birth clouds are resolved, the measured gas surface density remains constant even as resolution increases. There will be a transition between these two regimes,  as outlined in the diagram in Figure \ref{fig:diagram_resolution_dependence_gas_cloud}. The relation naturally becomes more complicated once the effect of multiple, sometimes overlapping birth clouds is taken into account. In particular, the flux dilution effect in the low resolution regime is reduced when contributions of many overlapping clouds are considered. 

We can measure the resolution dependence of $\Sigma_{\textrm{gas}}^{\textrm{BC}}$ by isolating its contribution to the total attenuation. We calculate it as: 
\begin{equation}
    \overline{\Sigma}_{\textrm{gas}}^{\textrm{BC}} = \overline{\Sigma}_{\textrm{gas}}^{\textrm{H}\alpha} - \overline{\Sigma}_{\textrm{gas}} 
    \label{eq:gas_correction}
\end{equation}
where $\overline{\Sigma}_{\textrm{gas}}^{\textrm{H}\alpha}$ is the H$\alpha$-weighted mean gas surface density and $\overline{\Sigma}_{\textrm{gas}}$ is the mean gas surface density in a galaxy. We use H$\alpha$ flux as the weight, allowing us to measure the gas surface density along the line of sight of star forming regions, as those dominate the H$\alpha$ emission. From this we subtract the mean gas surface density, which represents the contribution of the ambient ISM $\Sigma_{\textrm{gas}}^{\textrm{ISM}}$ to the total dust attenuation. 

\begin{figure}
    \centering
    \includegraphics[width=\columnwidth]{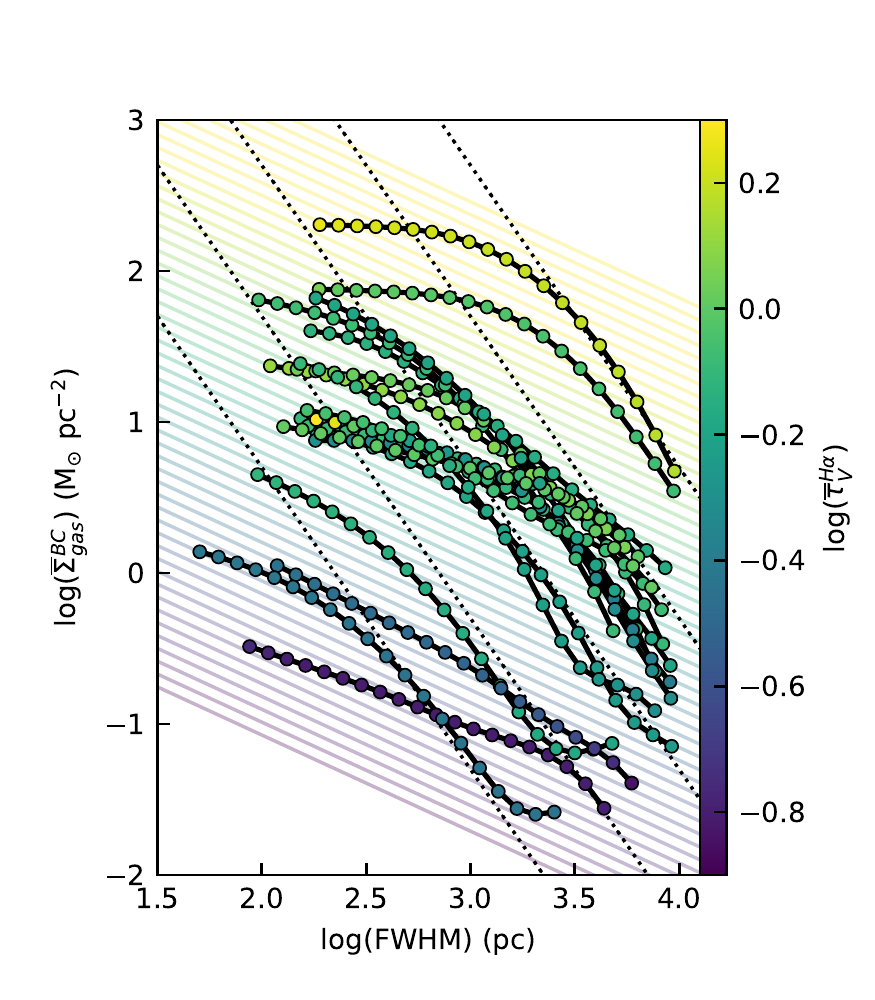}
    \caption{The measured $\overline{\Sigma}_{\textrm{gas}}^{\textrm{BC}}$ of the galaxies covered in both the PHANGS-ALMA and PHANGS-MUSE surveys at resolutions between $\sim2"$ and $\sim100"$ ($\sim$ 100 to 10000 pc). Optical depths are derived using the Balmer decrement from the PHANGS-MUSE data and molecular Hydrogen surface densities are derived from PHANGS-ALMA. The expected scaling relation for individual unresolved (clusters of) birth clouds ($\Sigma_{\textrm{gas}}^{\textrm{BC}} \propto \textrm{FWHM}^{-2}$) is shown with black dotted lines. The fitted relation for $\Sigma_{\textrm{gas}}^{\textrm{BC}}$ as a function of FWHM and $\tau_V$ is shown in the coloured lines where the colours represent different values of $\tau_V$ as specified by the colour bar.}
    \label{fig:Excess_SD_dependence_resolution_tau}
\end{figure}

We quantify the resolution dependence of $\Sigma_{\textrm{gas}}^{\textrm{BC}}$ using the molecular gas surface density maps from the CO(2-1) PHANGS-ALMA observations,  and the H$\alpha$ maps from the PHANGS-MUSE observations, for all 19 galaxies in the overlap of the two surveys. We do this by convolving the ALMA and MUSE maps to a range of resolutions between $\sim2"$ and $\sim100"$ ($\sim$ 100 to 10000 pc). Figure \ref{fig:Excess_SD_dependence_resolution_tau} shows $\overline{\Sigma}_{\textrm{gas}}^{\textrm{BC}}$ as defined in Eq. \ref{eq:gas_correction} for each galaxy, at each spatial resolution probed. We use the distances as in \cite{leroy2021} to derive the physical resolution of the observations. As expected, $\overline{\Sigma}_{\textrm{gas}}^{\textrm{BC}}$ varies with the physical scale, but also displays systematic galaxy-to-galaxy variations depending on the global optical depth, as high $\tau_{V}$ galaxies have more gas rich star forming regions compared to low $\tau_{V}$ galaxies. Here, the optical depth measurement is derived using the Balmer decrement and Equation \ref{eq:optical_depth_BD}. 

At the lowest resolutions (FWHM $\gtrsim 1$ kpc), Figure  \ref{fig:Excess_SD_dependence_resolution_tau} shows that most galaxies approach the scaling of FWHM$^{-2}$ (dotted lines), as expected. At high resolutions (FWHM $< 1$~kpc), the slope of the relation becomes more shallow, even reaching the theoretical constant scaling for the galaxies with the highest ISM densities (i.e. those with the largest global values of $\tau_{V}$). The transition between the two regimes (as described in the diagram in Fig. \ref{fig:diagram_resolution_dependence_gas_cloud}) occurs at different resolutions for different galaxies, depending on the physical scale of (clusters of) birth clouds. Due to the combined effects of many birth clouds blending together the resolution dependence of most galaxies is somewhere between the FWHM$^{-2}$ and constant dependencies, for most of the resolutions covered. 

\begin{figure}
    \centering
    \includegraphics[width=\columnwidth]{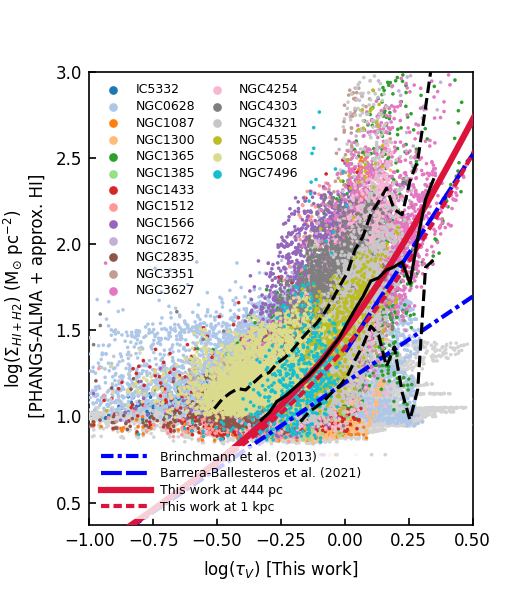}
    \caption{The relation between optical depth and gas surface density measured at 7.5" using PHANGS-ALMA to derive the molecular gas content and an assumed constant atomic hydrogen surface density of 6 M$_{\odot}$ pc$^{-2}$ \citep{leroy2021}. The data are coloured by each galaxy, as shown in the legend. Red lines show the resolved gas surface density relation we derived in Equations \ref{eq:resolved_gas_equation} and \ref{eq:gas_equation_correction} at the 444 pc which is the median physical resolution in the PHANGS sample (red, solid),  and at 1 kpc (red, short-dashed). The black solid and dashed lines show the running median and 16th and 84th percentile measurements. In blue we show the integrated gas surface density relation as in Equation \ref{eq:gas_equation_brinchmann} \citep{brinchmann2013} (dash-dotted) and the resolved relation at $\sim$1 kpc scale measured using EDGE-CALIFA \citep{barrera-ballesteros2021}.}
    \label{fig:MUSE_optical_depth_comparison_PHANGS_ALMA}
\end{figure}

Using the dependencies of $\Sigma_{\textrm{gas}}^{\textrm{BC}}$ on FWHM (in pc) and $\tau_V$ we construct a relation that describes the observed birth cloud gas surface density:
\begin{equation}
\label{eq:gas_equation_correction}
    \overline{\Sigma}_{\textrm{gas}}^{\textrm{BC}} = \frac{9.56\textrm{FWHM}^{-0.637}\tau_{V}^{2.98}}{\xi Z} (M_{\odot}\textrm{ pc}^{-2}).
\end{equation}
which is based on the best-fit linear relation in log-space. The parameters $\xi Z$ by definition represent the dust-to-gas ratio \citep{brinchmann2013} and are included to convert the dust based measurements of the optical depth to gas measurements but not considered as free parameters. The best fit relation is shown in Figure \ref{fig:Excess_SD_dependence_resolution_tau} with solid coloured lines with colours as in the colour bar. Clearly, a best-fitting model with a constant linear slope is an approximation, and cannot fully capture the two different regimes described above and the transition between them. With a larger sample, it would be possible in the future to fit a physically-motivated model that takes into account the expected transition from a slope of $\sim 0$ at high resolution, to $\sim -2$ at low resolution. 

The total resolved gas surface density relation can now be derived by adding the $\Sigma_{\textrm{gas}}^{\textrm{ISM}}$ and $\Sigma_{\textrm{gas}}^{\textrm{BC}}$ components together:
\begin{equation}
\label{eq:resolved_gas_equation}
\begin{split}
\overline{\Sigma}_{\textrm{gas}}^{\textrm{resolved}} & = \overline{\Sigma}_{\textrm{gas}}^{\textrm{ISM}}(\tau_{V}, \xi, Z) + \overline{\Sigma}_{\textrm{gas}}^{\textrm{BC}}(\textrm{FWHM}, \tau_{V}, \xi, Z) \\
 & =  \frac{0.2\tau_{V}}{\xi Z} + \frac{9.56\textrm{FWHM}^{-0.637}\tau_{V}^{2.98}}{\xi Z} (M_{\odot}\textrm{ pc}^{-2}), 
\end{split}
\end{equation}
where the ambient ISM ($\Sigma_{\textrm{gas}}^{\textrm{ISM}}$) and birth cloud ($\Sigma_{\textrm{gas}}^{\textrm{BC}}$) terms are given by Equations \ref{eq:gas_equation_brinchmann} and \ref{eq:resolved_gas_equation}, respectively.

Equipped with this resolution-dependent relation between gas surface density and optical depth/dust attenuation, we can now test how well gas surface densities can be estimated from optical emission lines flux measurements and photoionization modeling. We apply our photoionization models to the PHANGS-MUSE observations at $7.5"$ resolution. We train a simulation based inference model for the MUSE observations as the MUSE spectrograph does not cover the [OII]3727\AA\ and [OII]3729\AA\ emission lines. Due to these missing lines we have poor constraints on the ionization parameter measurements, however, the other parameters which directly feed into the gas surface density measurements are not affected. We compare the gas surface density measurements we derive using Equation \ref{eq:resolved_gas_equation} to those directly measured. As the optical depth measurement is sensitive to the total atomic$+$molecular gas content along the line of sight, we add to the molecular gas surface density measured from the PHANGS-ALMA CO(2-1) maps \citep{leroy2021} a constant atomic hydrogen surface density of 6 M$_{\odot}$ pc$^{-2}$ \citep{barrera-ballesteros2020}. We only apply our photoionization models to spaxels which have H$\alpha$ signal-to-noise > 15 and where the photoionization is caused by star formation activity, as determined by the \cite{kauffmann2003} criterion. 

In Figure \ref{fig:MUSE_optical_depth_comparison_PHANGS_ALMA} the solid red line shows the resolved $\tau_V-\Sigma_{\textrm{gas}}$ relation at 444 pc (median spatial scale in the sample at 7.5$"$) using Equation \ref{eq:resolved_gas_equation}. The black solid and dashed lines show the running median and 16th and 84th percentile measurements. The dash-dotted blue line shows the global relation from Equation \ref{eq:gas_equation_brinchmann} as defined by \cite{brinchmann2013}. The long-dashed blue line shows the measured relation between dust attenuation and molecular gas surface density at kpc scale in EDGE-CALIFA \citep[][]{barrera-ballesteros2021}, we plot this relation in the range of parameter space where molecular gas is the dominant component in the gas surface density. The comparison between the measured relation by \cite{barrera-ballesteros2021} to our parameterization at kpc scales (red, short-dashed) shows that these two relations are compatible. This figure clearly shows the importance of taking into account the resolution dependence of the relation between optical depth and gas surface density. 

\begin{figure}
    \centering
    \includegraphics[width=\columnwidth]{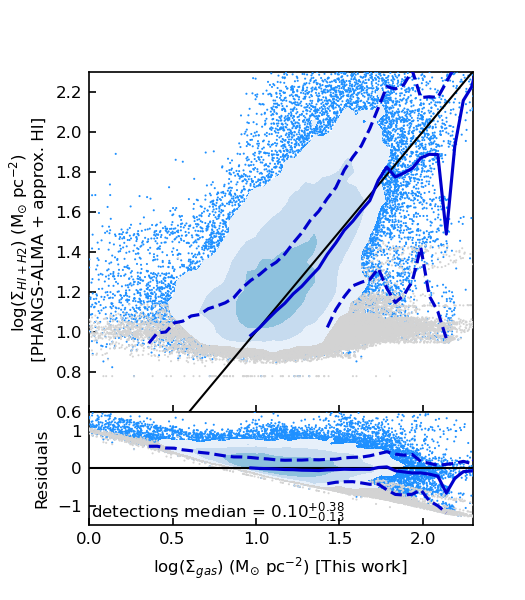}
    \caption{The relation between gas surface densities we derived from the optical emission and direct measurments using PHANGS-ALMA to derive the molecular gas content and an assumed constant atomic hydrogen surface density of 6 M$_{\odot}$ pc$^{-2}$ \citep{leroy2021}. The PHANGS-ALMA detections are shown in light blue (contours and data points) and the upper limits are shown in grey. The dark blue solid and dashed lines show the running median and 16th and 84th percentile measurements. The one-to-one relation is shown in black.}
    \label{fig:MUSE_gas_surface_density_comparison_PHANGS_ALMA}
\end{figure}

In Figure \ref{fig:MUSE_gas_surface_density_comparison_PHANGS_ALMA} we show a comparison between the gas surface densities we derive using the resolved gas surface density relation and the direct CO-based measurements from PHANGS-ALMA. Detections are shown in light blue (contours and data points), 3$\sigma$ upper limit measurements are shown in grey. The dark blue solid and dashed lines show the running median and 16th and 84th percentile measurements. The one-to-one relation is shown in black. These results show that the running median measurements follow the one-to-one relation very closely, and therefore that the resolution dependence is taken into account through Equation \ref{eq:gas_equation_correction} and \ref{eq:resolved_gas_equation}, optical depth is a good proxy for gas surface density, even in resolved observations. 

The scale-dependent nature of the $\tau_V - \Sigma_{\textrm{gas}}$ relation may be an important contributor to some of the disagreements reported in the literature when deriving cold gas mass surface densities from optical IFU emission line measurements using a calibration validated on integrated measurements \cite[e.g.][]{barrera-ballesteros2020, barrera-ballesteros2021}. The resolution dependence of the  $\tau_V - \Sigma_{\textrm{gas}}$ relation arises from the correlated distribution of gas clouds and star forming regions on small scales. This correlated distribution of star forming regions is also shown through the factor of $\sim$2 lower dust attenuation derived from stellar reddening compared to reddening of Balmer emission lines found by \cite{kreckel2013} which is attributed to the preferential location of star forming regions in gas and dust rich regions \citep{calzetti1994, calzetti2000, charlot2000, hao2011}.

The resolution dependence of measured ISM parameters is also demonstrated by \cite{grasha2022}. They show that low resolution measurements of ionization parameter and ISM pressure typically result in higher measured values than when observed at  high resolutions. This result is in agreement with our results, by extension of the result that just like optical depth, the integrated ionization parameter and ISM pressure measurements are biased towards H$\alpha$-bright regions. 

\subsubsection{Integrated gas mass comparisons to xCOLD GASS}
\label{sec:comparison_xcoldgass}

\begin{figure*}
    \centering
    \includegraphics[width=\textwidth]{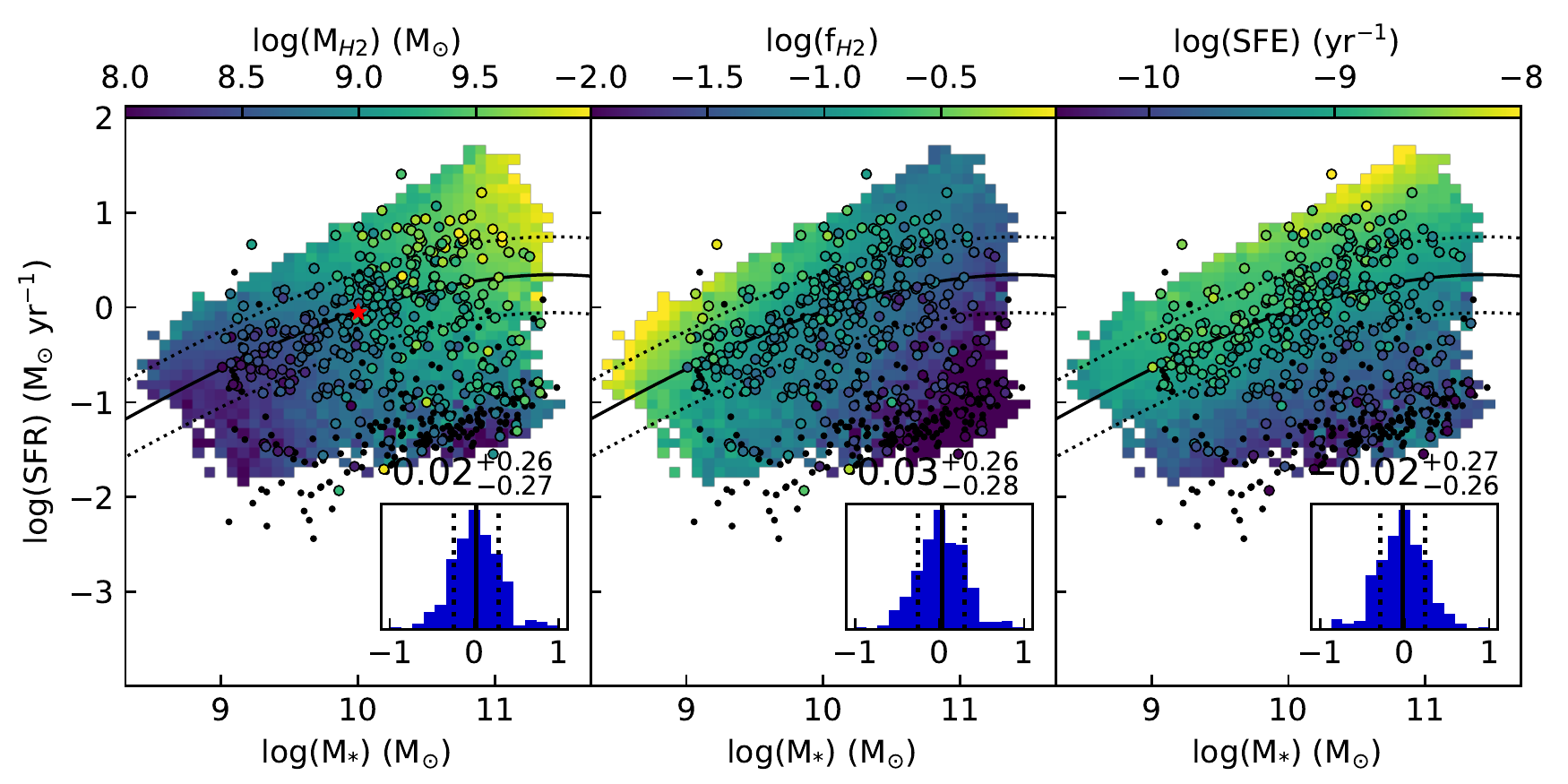}
    \caption{Molecular gas masses, gas fractions and star formation efficiencies derived from SDSS spectroscopy compared to CO-based measurements from the xCOLD GASS survey \citep{saintonge2017}. The binned data in the $M_{*}$-SFR plane are the median measurements derived in each bin with 10 measurements or more. The coloured data points show the xCOLD GASS measurements. The colour of each of these measurements shows the value of the molecular gas mass, gas fraction and star formation effiencies as described in the colour bar above each panel. The inset figures show the distribution of the xCOLD GASS measurement minus the median SDSS derived measurement in the corresponding bin in the $M_{*}$-SFR plane. The red star marks the galaxy for which we show the posterior distribution in Figure \ref{fig:corner_plot_example}.}
    \label{fig:MH2_FH2_SFE_comparison_xCOLDGASS}
\end{figure*}

We now test whether we can accurately infer the total molecular gas mass of galaxies using optical emission line information obtained from fibre observations with limited aperture sizes (e.g. 3\arcsec\ for SDSS and 1\arcsec\ for DESI). We use the SDSS sample described in Sec. \ref{sec:model_validation} for which we derive total molecular gas masses as:
\begin{equation}
    \label{eq:gas_surface_density_to_mass}
    \textrm{M}_{\textrm{H2}} = \frac{2\pi A}{e} \Sigma_{\textrm{gas}} R_{50}^{2} (M_{\odot}), 
\end{equation}
where $R_{50}$ are the r-band Petrosian half-light radii in pc from the NASA-Sloan Atlas \citep{blanton2011}, and $\Sigma_{\textrm{gas}}$ is the gas surface density as calculated through Equation \ref{eq:resolved_gas_equation}. The factor $\frac{2\pi}{e}$ results from the integration of the assumed geometry of the gas in an exponential disk. The factor $A$ accounts for the combined effects of several factors: (1) the relation between the measured $\Sigma_{\textrm{gas}}$ and the gas surface density at the effective radius of a Sersic profile, (2) the effect of the average inclination of a galaxy and (3) the relation between the r-band $R_{50}$ and the effective radius of the gas profile \citep{casasola2017}. 

To calibrate the value of the factor $A$, we use the total molecular gas masses of galaxies in the xCOLD GASS survey, which are measured using CO(1-0) line luminosities and a metallicity dependent $\alpha_{CO}$ conversion function \citep{saintonge2017}. We do not make direct comparisons between the SDSS- and xCOLD GASS-derived $M_{\textrm{H2}}$ values, as the low redshifts of the xCOLD GASS sample ($z<0.05$) mean that aperture effects are significant, and that the important [OII]3627 and [OII]3629 emission lines are not always accessible in the SDSS spectra. Instead, we compare the molecular gas masses of the xCOLD GASS galaxies with the median value in our SDSS sample at fixed stellar mass and star formation rate. Using this method, we calibrate a value of $\log(A) = 0.15$. We have not included atomic Hydrogen in the calibration, as a large fraction of the atomic Hydrogen of galaxies is located in an extended region around the galaxy which is not probed by our measurements using emission line spectroscopy.

In Figure \ref{fig:MH2_FH2_SFE_comparison_xCOLDGASS}, we show the distribution of the molecular gas masses, gas fractions ($f_{H2} \equiv M_{H2}/M_{\ast}$) and star formation efficiencies ($SFE \equiv SFR / M_{H2}$) we derive for the SDSS sample across the $M_{*}$-SFR plane. Overplotted are the xCOLD GASS galaxies with the same measurements inferred from direct CO(1-0) observations.  The figure shows that we are able to recover gas measurements for galaxies with a wide range of stellar masses and star formation rates, with the method performing well for star-forming main sequence galaxies and massive passive objects. No trends are visible in the residuals as a function of star formation rate or stellar mass at a level comparable to or greater than the typical measurement uncertainties. The median scatter of $\sim0.25-0.3$ dex is similar to other calibrations of SDSS-based molecular gas masses, such as \cite{concas2019}, \cite{yesuf2019} and \cite{piotrowska2020}. An advantage of our method is that by calibrating on binned measurements in the $M_{*}$-SFR plane, we avoid some limitations due to SDSS aperture effects which are largest at the low redshifts of the xCOLD GASS survey. 

It is important to note that whilst there is a radial gas surface density gradient in galaxies, as assumed for example by Equation \ref{eq:gas_surface_density_to_mass}, this does not directly affect the measured gas surface density through optical emission lines within an aperture. Because integrated optical emission line measurements are dominated by the brightest sub-regions, areas with lower surface densities such as outer disc regions do not contribute significantly. This means that the measured optical depth is not strongly dependent on the aperture size. The other aperture dependent effect that could affect the gas mass measurements would be metallicity gradients. However, as shown by \cite{lutz2021}, there are no strong systematic trends that allow us to predict the radial metallicity profile of any galaxy, based on its other properties. Attempting to statistically-correct for metallicity gradients within the aperture would therefore likely only introduce noise, we choose to assume flat metallicity profiles.

\subsection{Understanding the scatter of the mass-metallicity relation}
\label{sec:MZR}

\begin{figure*}
    \centering
    \includegraphics[width=2\columnwidth]{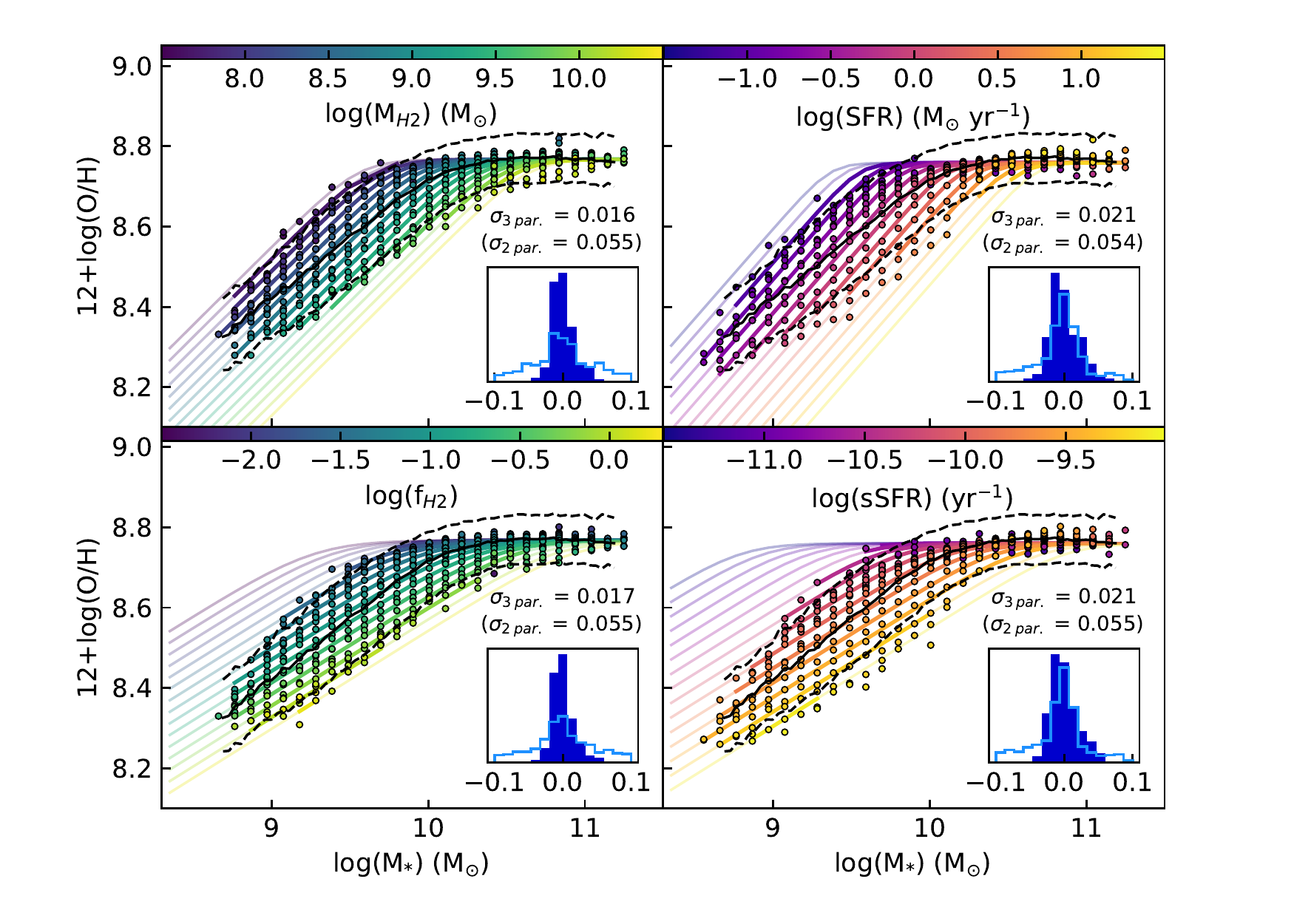}
    \caption{The scatter in the MZR as a function of molecular gas mass (top-left), molecular gas fraction (bottom-left), star formation rate(top-right) and specific star formation rate(bottom-right). The data points show the mass metallicity relation binned along stellar mass and the 3$^{\textrm{rd}}$ parameters listed above, coloured by the 3$^{\textrm{rd}}$ parameter value. The coloured lines show the best fit relation according to equation \ref{eq:mass_metallicity_scatter_function}. The black solid and dotted lines show the median and 16$^{\textrm{th}}$ and 84$^{\textrm{th}}$ percentiles of the MZR. The inset plots show the residuals between the binned data and the best fit 3-parameter relation (darkblue) and the MZR (lightblue).}
    \label{fig:MZR_binned}
\end{figure*}

We use the SDSS DR8 sample to study the comparative roles of gas and star formation in the scatter of the MZR, the relation between stellar mass and gas-phase metallicity. We use the same sample of galaxies as used in Section \ref{sec:comparison_xcoldgass}, but exclude galaxies with star formation rate and stellar mass aperture correction factors smaller than 0.1 or larger than 1.0, spectra flagged unreliable and spectra with non-galaxy target types. The stellar mass measurements are from the MPA-JHU catalog \citep{kauffmann2003a}. We use the \cite{pettini2004} O3N2 indicator to calculate the metallicities that go in our MZR; using the metallicities from the photoionization modelling, in addition to the gas surface densities also derived from the models, could introduce spurious correlations due to parameter degeneracies, as has been shown by \cite{mingozzi2020} for metallicity and ionization parameter measurements.

To explore the nature of the scatter, we fit the MZR using the functional form defined in \cite{curti2020}, which includes the contribution of a third parameter, traditionally taken to be either SFR or $M_{\textrm{gas}}$:
\begin{equation}
\label{eq:mass_metallicity_scatter_function}
    Z(M, A) = Z_0 - \frac{\gamma}{\beta} \mathrm{log}(1 + (M/M_0(A))^{-\beta} ) .
\end{equation}
In this equation, $Z$ is the metallicity, $M$ the stellar mass, $Z_0$ the metallicity at the high mass end, $\gamma$ the power law index of the slope at low stellar masses, $\beta$ constrains the width of the transition region, $M_0(A)$ the turn-over stellar mass (where $\textrm{log}(M_0(A)) = m_0 + m_1\textrm{log}(A)$) and $A$ is the third parameter used to describe the scatter in the relation.

\begin{figure*}
    \centering
    \includegraphics[width=2\columnwidth]{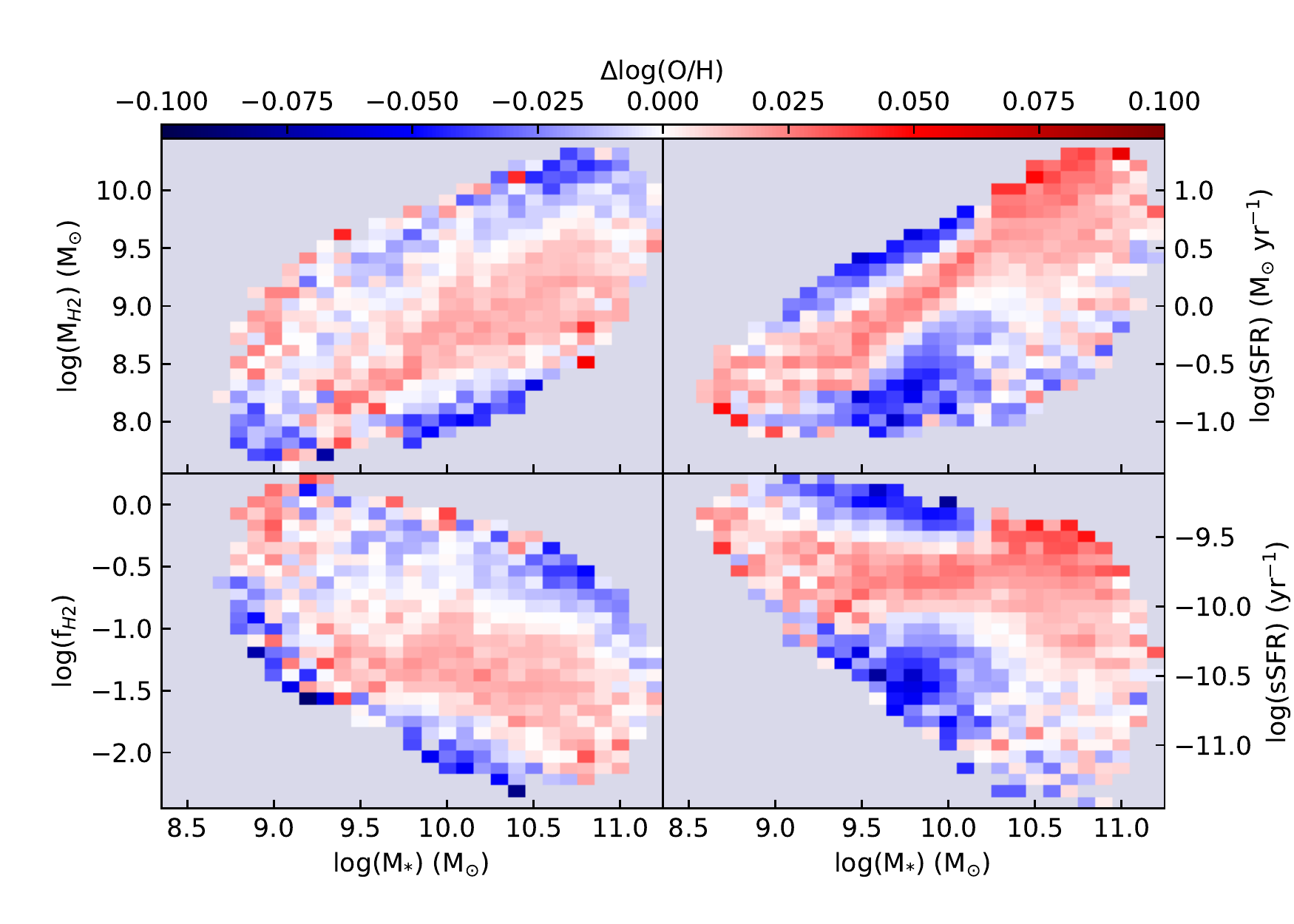}
    \caption{The residuals in the 3-parameter relation between stellar mass, metallicity and gas surface density (top-left), gas fraction (bottom-left), star formation rate(top-right) and specific star formation rate(bottom-right). Only bins with 25 or more measurements are included.}
    \label{fig:MZR_binned_metallicity_residuals}
\end{figure*}

In our comparisons we use 4 quantities to derive the scatter in the MZR: molecular gas mass, molecular gas fraction ($F_{H2} = M_{H2}/M_{*}$), star formation rate and specific star formation rate. We use our calibrated gas mass measurements, and the aperture-corrected star formation rates from \cite{brinchmann2004}. We fit the relation after grouping the data into 30 bins along stellar mass and each of the 3$^{\textrm{rd}}$ parameters we use. Both the binned data points and best-fit models are shown in Fig. \ref{fig:MZR_binned}, with the values of the best fit parameters presented in Table \ref{tab:mzr_best_fits}. The introduction of the 3$^{\textrm{rd}}$ parameters reduces the scatter in each of the chosen cases. In Figure \ref{fig:MZR_binned} the residuals of the 3 parameter relation are given by $\sigma_{3\: par.}$ and the residuals in the MZR given this particular binning by $\sigma_{2\: par.}$. 
A large value of $\sigma_{2\: par.}$ shows that the chosen 3$^{\textrm{rd}}$ parameter effectively describes the scatter in the MZR. The final residual $\sigma_{3\: par.}$ describes how well this 3-parameter relation reduces the scatter after fitting the binned data with the function in Equation \ref{eq:mass_metallicity_scatter_function}. Figure \ref{fig:MZR_binned} shows that each of these quantities describes the scatter in the MZR by a comparable amount. The molecular gas mass and gas fraction parameters reduce the scatter in the 3-parameter relation the most with a final residual scatter of $\sim$0.016 dex. We also show the distribution of the metallicity residuals over the individual bins in Figure \ref{fig:MZR_binned_metallicity_residuals}. 

The role of a third parameter in driving the scatter of the MZR is an area which has been studied extensively \citep[e.g.][]{ellison2008}. The effect of star formation in the scatter of the MZR was formalized into the fundamental metallicity relation (FMR) by \cite{mannucci2010}: the relation between stellar mass, star formation and metallicity. Further studies of the FMR expand the relation to lower stellar masses and higher redshifts \citep{mannucci2011, cresci2019, curti2020}. However, gas has also been shown to play an important role in the scatter of the MZR by \cite{brinchmann2013, bothwell2013, bothwell2016, lara-lopez2013, hughes2013} and \cite{brown2018} through direct measurements of atomic and molecular gas content. Their results suggest that gas is more fundamental than SFR in  driving the scatter in the MZR. The respective roles of star formation and gas have also been studied using the EAGLE simulations \citep{loon2021}. They show that gas fraction correlates most strongly with the scatter in the MZR. Our measurements show that gas and star formation processes both describe the scatter in the MZR, but that molecular gas mass and molecular gas fraction are most effective at reducing the scatter, in agreement with similar studies performed with HI and CO observations and the predictions of simulations and theory. 
This result importantly shows that cold gas masses derived from emission line measurements are sufficiently accurate to derive the role of cold gas in galaxy evolution processes. 

\begin{table}
 \centering
 \caption{Best-fit parameter values for the mass-metallicity + 3$^{\textrm{rd}}$ parameter relations using the parametrization from Equation \ref{eq:mass_metallicity_scatter_function}.}
 \label{tab:mzr_best_fits}
\begin{tabular}{llllll}
\hline
\hline
\textbf{Third parameter}      & \textbf{$Z_0$} & \textbf{$m_0$} & \textbf{$m_1$} & \textbf{$\gamma$} & \textbf{$\beta$} \\ \hline
---                           & 8.77           & 10.14          & ---            & 0.31              & 2.33             \\
Molecular gas mass            & 8.77           & 6.74           & 0.37           & 0.39              & 3.61             \\
Molecular gas fraction        & 8.77           & 10.77          & 0.61           & 0.24              & 2.17             \\
star formation rate           & 8.76           & 10.02          & 0.45           & 0.45              & 5.73             \\
specific star formation rate  & 8.76           & 18.69          & 0.86           & 0.24              & 2.84             \\
\hline
\end{tabular}
\end{table}

\section{Summary and conclusion}
\label{sec:conclusion}
We use photoionization models and Simulation Based Inference to derive cold interstellar medium masses of galaxies from optical emission lines. The methods we use in our photoionization models are a combination of successful features from previous work. We mainly base our models on the method used by \cite{byler2017} whilst reintroducing the dust-to-metal ($\xi$) parameter from the \cite{charlot2001} models which is essential to derive gas masses from the photoionization models \citep{brinchmann2013}. The implementation of Simulation Based Inference opens the door to applying photoionization models to the largest samples of galaxies. Inference using MCMC methods demand significantly more computing power and inference methods sparsely exploring the parameter space lose much of the information on degeneracies between parameters. This makes Simulation Based Inference a crucial tool to make photoionization modelling possible for the samples sizes of the coming era of large spectroscopic surveys, which will deliver tens of millions spectroscopic observations of galaxies.

To validate our results we compare our measurement to a range of other measurements for many of the quantities we derive from the photoionization modelling. We provide validation of our gas mass measurements for both resolved and fully integrated observations. To do so we derive a resolution-dependent relation between gas surface density and optical depth. This makes $\tau_{V}$-$\Sigma_{\textrm{gas}}$ relations applicable over a wide range of resolutions. We demonstrate this by deriving gas surface densities from the optical depth measurements of galaxies in the PHANGS-MUSE survey at resolutions of $\sim7.5"$ ($\sim$500 pc). We also show that we can derive gas masses for galaxies observed in the SDSS survey, which we validated using xCOLD GASS direct measurements. Using this technique we can accurately derive the molecular gas mass of galaxies with a wide range of stellar masses and star formation rates. 

In this work we show that the gas mass measurements we derive allow us to study the role of gas in galaxy evolutionary processes. This is important to establish as the measurement of gas masses of galaxies using the Balmer decrement and other optical emission lines is a less precise measurement method than other available methods, however, the sample sizes accessed are unmatched. We demonstrate sufficient accuracy by studying the role of molecular gas in the mass-metallicity relation. We confirm that gas plays a fundamental role in the scatter of the relation, which has been shown previously through direct atomic and molecular gas measurements and simulations \citep[e.g.][]{bothwell2013, bothwell2016, loon2021}. This result demonstrates that cold gas masses derived from emission line measurements are sufficiently accurate to derive the role of cold gas in galaxy evolution, opening new parameter space with large, deep next-generation surveys such as DESI.

\section*{Acknowledgements}
We thank the referee, Brent Groves, for extremely constructive feedback. The work has also benefited from discussions with Mike Barlow, Serena Viti, Sara Ellison and Tim Heckman. 

Besides software packages already mentioned in the main body of this paper, this work has also made use of Python\footnote{https://www.python.org/} and the Python packages: astropy\footnote{https://www.astropy.org/} \citep{astropy2013}, NumPy\footnote{https://numpy.org/}, matplotlib\footnote{https://matplotlib.org/} \citep{hunter2007}, pandas\footnote{https://pandas.pydata.org/} \citep{reback2020} and SciPy\footnote{https://scipy.org/}.

Funding for SDSS-III has been provided by the Alfred P. Sloan Foundation, the Participating Institutions, the National Science Foundation, and the U.S. Department of Energy Office of Science. The SDSS-III web site is http://www.sdss3.org/. SDSS-III is managed by the Astrophysical Research Consortium for the Participating Institutions of the SDSS-III Collaboration including the University of Arizona, the Brazilian Participation Group, Brookhaven National Laboratory, University of Cambridge, Carnegie Mellon University, University of Florida, the French Participation Group, the German Participation Group, Harvard University, the Instituto de Astrofisica de Canarias, the Michigan State/Notre Dame/JINA Participation Group, Johns Hopkins University, Lawrence Berkeley National Laboratory, Max Planck Institute for Astrophysics, Max Planck Institute for Extraterrestrial Physics, New Mexico State University, New York University, Ohio State University, Pennsylvania State University, University of Portsmouth, Princeton University, the Spanish Participation Group, University of Tokyo, University of Utah, Vanderbilt University, University of Virginia, University of Washington, and Yale University.  

\section*{Data Availability}

The data underlying this article were accessed from the 8th data release of the SDSS survey\footnote{https://www.sdss3.org/dr8/} \citep{abazajian2009}, the PHANGS survey\footnote{http://phangs.org} \citep{emsellem2021,leroy2021} and the xCOLD GASS survey\footnote{http://www.star.ucl.ac.uk/xCOLDGASS/} \citep{saintonge2017}. The grids of photoionization models generated for this article are available at \href{https://github.com/dirkscholte/paper_spectroscopic_gas_masses}{https://github.com/dirkscholte/paper\_spectroscopic\_gas\_masses}. The derived data generated in this research will be shared on reasonable request to the corresponding author.



\bibliographystyle{mnras}
\bibliography{references} 




\appendix

\section{Comparisons to other metallicity calibrations}
\label{sec:appendix_metallicities}

We compare our gas phase metallicities to a range of metallicity calibrations using common strong line metallicity indicators. For the O3N2 and N2 calibrators we show parametrizations by \cite{pettini2004, nagao2006, maiolino2008, marino2013, brown2016} and \cite{curti2017}. The O3N2 indicator is described in Equation \ref{eq:O3N2_indicator} and the N2 indicator is described by \citep{storchi-bergmann1994}:
\begin{equation}
    \textrm{N2 = log}\left(\frac{\textrm{[NII]}6584}{\textrm{H}\alpha}\right).
\end{equation}
We also use the R23 indicator \citep{pagel1979}:
\begin{equation}
    \textrm{R23 = log}\left(\frac{\textrm{[OII]}3726, \textrm{[OII]}3729 + \textrm{[OIII]}4959, \textrm{[OIII]}5007}{\textrm{H}\beta}\right)
\end{equation}
with parametrizations from \cite{mcgaugh1991, zaritsky1994, nagao2006, maiolino2008} and \cite{curti2017}. And the N2O2 indicator \citep{alloin1979}:
\footnotetext[1]{Dust attenuation corrected using the Balmer decrement and equations \ref{CH2000_transmission} and \ref{dust_attenuation}. Assuming an intrinsic Balmer decrement of 2.86.}
\begin{equation}
    \textrm{N2O2 = log}\left(\frac{\textrm{[NII]}6584\textrm{\footnotemark[1]}}{\textrm{[OII]}3726, \textrm{[OII]}3729\textrm{\footnotemark[1]}}\right).
\end{equation}
with parametrizations from \cite{kewley2002, nagao2006} and \cite{brown2016}. For the N2O2 indicator the attenuation corrected line strengths have to be used. The attenuation correction is performed using the Balmer decrement. 

\begin{figure}
    \centering
    \includegraphics[width=0.95\columnwidth]{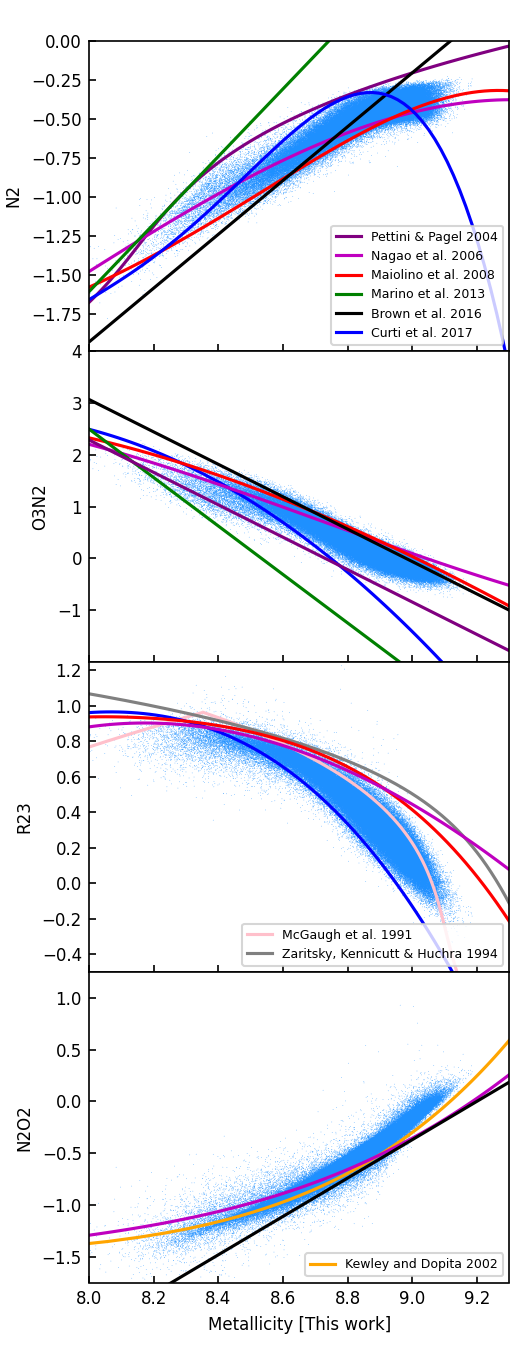}
    \caption{A comparison between the gas phase metallicities we derived and other calibrations through several commonly used metallicity indicators. In these figures the data points are coloured using the ionisation parameter with the same colourbar as in Figure \ref{fig:model_metallicity_vs_PP04_O3N2}.}
    \label{fig:metallicity_indicator_comparison}
\end{figure}

The comparisons in Figure \ref{fig:metallicity_indicator_comparison} show that the metallicities we derive fit within the range of previous calibrations. We note that our metallicity estimates are somewhat lower than typical theoretical calibrations using photoionization modelling.

\bsp	
\label{lastpage}
\end{document}